\documentclass[aps,prl,showpacs, twocolumn]{revtex4}
\usepackage{graphicx}
\usepackage{bm}
\usepackage{amsmath}
\usepackage{bbold}
\usepackage[utf8]{inputenc}

\providecommand{\bra}[1]{\langle #1 \rvert}
\providecommand{\ket}[1]{\lvert #1 \rangle}

\providecommand{\be}{\begin{equation}}
\providecommand{\ee}{\end{equation}}
\providecommand{\ba}{\begin{eqnarray}}
\providecommand{\ea}{\end{eqnarray}}

\begin{document}
\title{Finite temperature reservoir engineering and entanglement dynamics  }
 
\author{ S. Fedortchenko$^1$, A. Keller$^2$,   T. Coudreau$^1$, and P. Milman$^1$ }

\affiliation{$^{1}$Laboratoire Mat\' eriaux et Ph\' enom\`enes Quantiques, Sorbonne Paris Cit\' e, Universit\' e Paris Diderot, CNRS UMR 7162, 75013, Paris, France}
\affiliation{$^{2}$Univ. Paris-Sud 11, Institut des Sciences Mol\' eculaires d'Orsay (CNRS), B\^atiment 350--Campus d'Orsay, 91405 Orsay Cedex, France}

\begin{abstract}
We propose experimental methods to engineer reservoirs at arbitrary temperature which are feasible with current technology. Our results generalize to mixed states the possibility of  quantum state engineering through controlled decoherence. Finite temperature  engineered reservoirs can lead to the experimental observation of thermal entanglement --the appearance and increase of entanglement with temperature-- to the study of the dependence of finite time disentanglement and revival with temperature, quantum thermodynamical effects, among many other applications, enlarging the comprehension of temperature dependent entanglement properties. 
\end{abstract}
\vskip2pc

 
\vskip2pc 
\maketitle

For the vast majority of experimentally controllable and measurable quantum systems, interaction with the environment leads to a decoherence process that rapidly and irreversibly destroys the quantum properties of the studied system. Decoherence impedes the large scale applications of quantum mechanics, as quantum information and quantum metrology. Such unavoidable coupling between a system and an inaccessible environment  privileges so-called pointer states \cite{Zurek}. When interaction with the reservoir prevails over the system's free Hamiltonian, the specific properties of the system--environment coupling determine the steady, or more stable states of the considered system. Decoherence owes its bad reputation to the fact that, for experimentally relevant situations, these steady states usually do not display nonclassical properties. However, quantum reservoir engineering \cite{Poyatos, Carvalho, Wineland} showed that decoherence can be rendered compatible with the preservation of quantum properties. Indeed, by judiciously engineering the effects of exotic system--reservoir couplings, quantum states with useful quantum properties, such as entanglement or quantum coherence  \cite{Blatt, Wineland2, Krauter, SC, SCexp}, can turn out to be the steady states of the decoherence process.  Reservoir engineering can be used to protect interesting quantum states from decoherence, even in the case where a ``natural" reservoir is present \cite{Carvalho, Rouchon}. As a matter of fact, an engineered reservoir can dominate a system's dynamics, turning the total system's steady state arbitrarily close to its own steady state, that can be controllably chosen.

The experimental implementation of reservoir engineering  became achievable with the rapid technological development of quantum devices \cite{Haroche13, Wineland13}. Recent experimental results demonstrate the production of steady maximally entangled states through controlled dissipation in trapped ions \cite{Blatt, Wineland2} and superconducting systems \cite{SCexp}.  Most theoretical proposals and experimental realizations are focused on engineering zero temperature reservoirs, since they can lead to the production and protection of  pure states. For finite temperature, the effects of a local thermal reservoir in an initially prepared entangled state were studied in \cite{Steve}. 

The steady state of a finite temperature reservoir is mixed. Finite temperature reservoir engineering can lead to the production and protection of mixed states with interesting entanglement properties,  a problem that remains unexploited in the literature. Nevertheless, there are a number of interesting and unusual  entanglement properties that engineered thermal reservoir can help to reveal and fully exploit for applications. One example is thermal entangled states, states that are separable at zero temperature, but entangled at finite temperature \cite{Thermal1, Thermal2}. They are the eigenstates of a strongly interacting spin system coupled to a reservoir that, due to the strongly interacting Hamiltonian, is non-local, {\it i.e.} it acts in many spins at the same time. Experimental observation of thermal entanglement is possible in strongly interacting many body systems \cite{NMR, Ghosh, Vertesi, Brukner, Tatiana}. However, in such systems, entanglement cannot be extracted and used as a resource for quantum based protocols. Moreover,  the system-reservoir Hamiltonian is fixed by the specific material under study, and cannot be engineered and controlled. As a matter of fact, engineering system-reservoir couplings leading to the controllable production of thermal entanglement can provide a tool to better understand the complex structure of the many-body systems where it can be experimentally observed, as well as its coupling to an environment.

Another application of finite temperature reservoirs lies in the study of the behavior of entanglement in the presence of dissipation. It was shown in \cite{Marcelo, Eberly, SteveSD} that some two-qubit entangled states, in the presence of a zero temperature reservoir independently coupled to each qubit, completely disentangle at finite time. In \cite{James}, the case of finite temperature reservoirs was theoretically studied, and it was shown that almost all states display finite time disentanglement for finite temperature. However, no experimental evidence of this fact was provided so far, and finite temperature reservoir engineering is certainly an important tool towards this goal, as well as to the environment induced appearance of entanglement \cite{Birth}. Finally, quantum reservoir engineering at finite temperatures can find applications in simulating in more realistic scenarii of quantum transport problems \cite{Transp1, Transp2} and in providing an environment to the study of  quantum thermodynamics \cite{Thermo}.  The realization of finite temperature engineered reservoir will lead to the experimental evidencing of a broader collection of unusual quantum phenomena connecting  entanglement and temperature.

In the present contribution, we provide theoretical proposals leading to finite temperature reservoir engineering which are well adapted to the state-of-the art of different set-ups successfully used to  demonstrate quantum protocols. We detail these ideas for a trapped ion system and in \cite{SI} we show how to implement them using different modes of single photons. Extending the proposed ideas to other set-ups, as superconducting qubits, is also possible, but will not be detailed here \cite{SCexp}.

From an Hamiltonian describing the coupling between the system and an environment at temperature $T$ we can trace out the environment degrees of freedom in the Born-Markov limit  supposing that system and environment are uncorrelated at initial time $t=0$. This leads to a master equation ( Lindblad equation \cite{Lindblad}) governing the dynamics of the system's reduced density matrix, $\hat \rho$: 
\begin{eqnarray}
&\dot {\hat{\rho}}&  = \sum_{i=1}^N \Gamma \big(  \bar n_i  + 1 \big) \bigg(\hat c_{i} \hat{\rho} \hat c^{\dagger}_{i} - \frac{1}{2} \Big\lbrace  \hat c^{\dagger}_{i} \hat c_{i} , \hat{\rho} \Big\rbrace \bigg)    \nonumber \\
&&+\sum_{i=1}^N \Gamma \,  \bar n_i  \bigg(  \hat c^{\dagger}_{i} \hat{\rho}  \hat c_{i} - \frac{1}{2} \Big\lbrace \hat c_{i}  \hat c^{\dagger}_{i}, \hat{\rho}\Big\rbrace \bigg),
\label{MasterEq2}
\end{eqnarray} 
where $\hat \rho$ is composed by $N$ sub-systems that are independently coupled to the reservoir. In the present manuscript, we will consider that each subsystem is a  qubit. The operators $\hat c_i$, $\hat c_i^{\dagger}$ ,  also called {\it jump operators} \cite{Castin}, act independently on each qubit, and describe the action of the environment to the system. $\Gamma$ is a rate related to the system--reservoir coupling constant, and will be supposed to be the same for each sub-system.  $\bar n_i$ is a function of  the reservoir's temperature $T$: the reservoir can be interpreted as composed by an infinity of harmonic oscillators, and $\bar n_i$ is the average number of quanta in the mode resonantly coupled to the $i$-th qubit. In the general case, where qubits are encoded in physical systems with different characteristic frequencies, $\bar n_i$ can depend on $i$ to ensure that the reservoir is at constant temperature $T$. Notice that, for $T=0$, $\bar n_i=0$. 

The exact form of the coupling between the system and the environment determines operators $\hat c_i$($\hat c_i^{\dagger}$) and the steady state of Eq.~\eqref{MasterEq2}. For $T=0$, the steady states are the eigenstates of $\hat c_i$ with zero eigenvalue. For this reason, operators $\hat c_i$  play a central role in the decoherence process and the protection or disappearance of quantum properties \cite{Buch}.  

\begin{figure}[t!]
\centering
\includegraphics[width=0.49\textwidth]{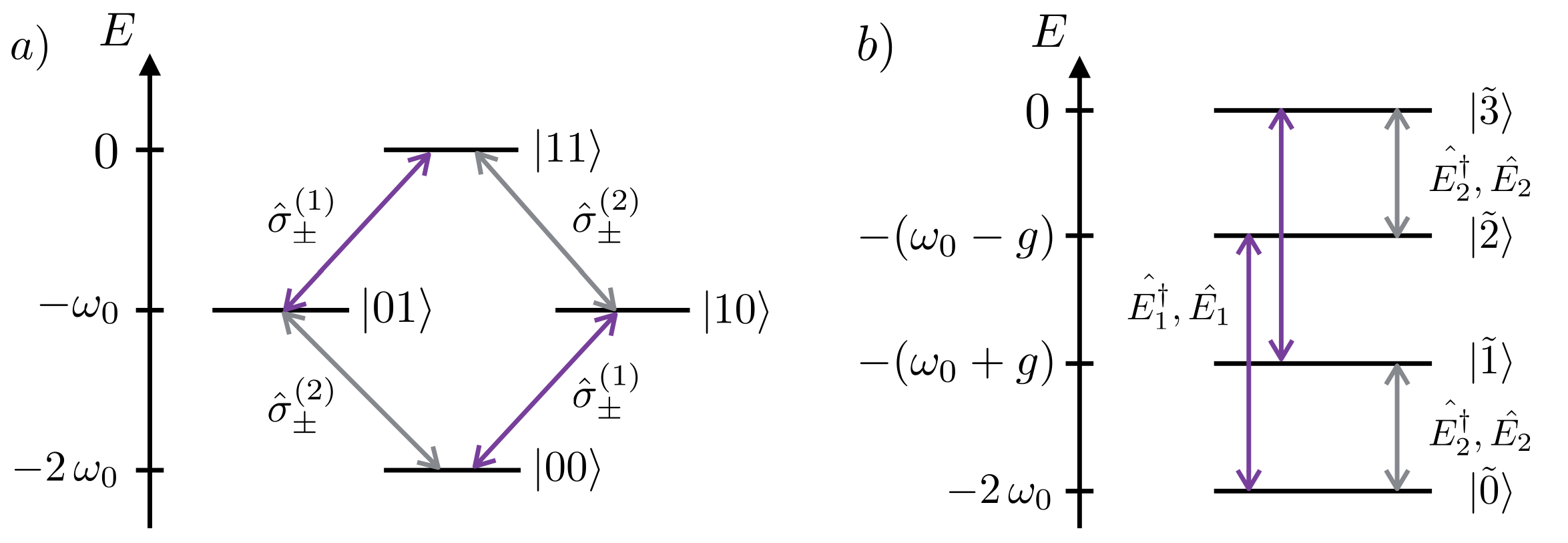}
\caption{ Level scheme and energy of states of a 2-qubit chain that coupled by jump operators. The arrows represent the possible transitions between the levels, governed by the jump operators. In (a) jump operators  are given by $\hat \sigma_{\pm}^{(1(2))}$. In(b) they are transformed by $\hat U$, leading to $\hat E_i= (  \hat{\sigma}_{-}^{(i)} + i \, \hat{\sigma}_{z}^{(i)} \hat{\sigma}_{-}^{(j)}   ) / \sqrt{2}$, $i,j=1,2$. }
\label{fig1}
\end{figure}

The considered qubits  are usually encoded on real or artificial atomic systems, as trapped atoms and ions or superconducting circuits.  In such systems, Eq. (\ref{MasterEq2}) is used to describe, for instance, the radiative atomic decay and absorption from the reservoir, and $\hat c_i=\hat \sigma_-^{(i)}=\hat \sigma_x^{(i)}-i \hat \sigma_y^{(i)} $. In this case, the steady state of the system is given by a thermal distribution, $\hat \rho_S=\frac{1}{Z} \sum_{k}^N e^{- E_{k} / k_{B} T} \vert k \rangle \langle k \vert$, where $Z$ is the partition function and states $\{ \vert k \rangle \}$ form the product state basis of $N$ qubits. $E_k$ is the energy associated to state $\ket{k}$. Such thermal distributions cure,fly have no application in quantum information. In Fig. \ref{fig1} (a), we show an example of the level scheme and energies of a system with $N=2$, together with the associated jump operators $\hat \sigma_-^{(i)}$ ($\hat \sigma_+^{(i)}$)  connecting states that differ by one excitation.   

If $\hat \rho_S$ is a stationary state of (\ref{MasterEq2}), we have that $\dot {\hat \rho}_S=0$, by definition. Thus, for any unitary transformation $\hat U$, defining $\tilde {\hat  \rho}_S= \hat U {\hat \rho}_S \hat U^{\dagger}$, $\hat U \dot {\hat \rho}_S \hat U^{\dagger}= \dot {\tilde {\hat \rho}}_S =0$. It follows immediately that  ${\tilde {\hat \rho}}_S $ is a stationary state of a transformed Lindbald equation, corresponding to applying to  (\ref{MasterEq2}) the substitution $\hat c_i  \rightarrow \hat E_{i}= \hat U \hat c_i \hat U^{\dagger}$ and $\hat c_i^\dagger  \rightarrow \hat E_i^{\dagger} =\hat U^{\dagger} \hat c_i^{\dagger} \hat U$. That is, in order to have ${\tilde {\hat \rho}}_S$ as a stationary state, one must simulate the non-unitary dynamics of a Lindblad equation where  $ \hat E_i$ is the jump operator instead of $ \hat c_i $.  Notice that while  $\hat c_i$ creates transitions (quantum jumps) between states $\ket{k}$ and $\ket{k'}$ of the product state basis, $ \hat E_i$ couples states $\hat U \ket{k}$ to $\hat U\ket{k'}$ (and analogously for $ \hat E_i^{\dagger}$). The transformed basis $\hat U \ket{k}$ to $\hat U\ket{k'}$ can be, for instance, formed by entangled states. 
 
An alternative description of the damping and decoherence process  is given by the generalized amplitude damping channel associated to Eq.~(\ref{MasterEq2}), that is,  the map ${\cal E}_{t}: \rho(t=0) \rightarrow \rho(t) = {\cal E}_{t}(\rho(t=0))$, that can be written in terms of the Kraus operators  $\hat M_j^{(i)}(t) $ as~:
\begin{equation}\label{Kraus}
{\cal E}_{t}(\hat \rho)=  \bigotimes_i^N \left ( \sum_j^4  \hat M_j^{(i)}(t) \hat \rho \hat M_j^{\dagger (i)}(t)\right ), 
\end{equation}
where $\hat M_j^{(i)}(t)$ is  $j$-th Kraus operator described in \cite{Chuang, Squeeze} acting on the $i$-th qubit.  Kraus operators are related to jump operators as follows: $\hat M_2^{(i)} \propto \hat c_i$ and  $\hat M_4^{(i)}\propto \hat c_i^{\dagger}$, while  $\hat M_1^{(i)}$ and $\hat M_3^{(i)}$ represent the system's evolution when  no quantum jumps occurs. The unitary transformation $\hat U$ undergone by the jump operators also applies to Kraus operators  : $\hat M_j^{(i)} \rightarrow \tilde {\hat M}_j^{(i)}=\hat U \hat M_j^{(i)} \hat U^{\dagger}$. Notice that when $\hat U$ is an entangling operation,  $\tilde {\hat M}^{(i)}_j$ will act on more than one qubit. In view of experimental implementations, the map (\ref{Kraus}) is a convenient approach, as we detail in the following.

\begin{figure}[t!]
\centering
\includegraphics[width=0.4\textwidth]{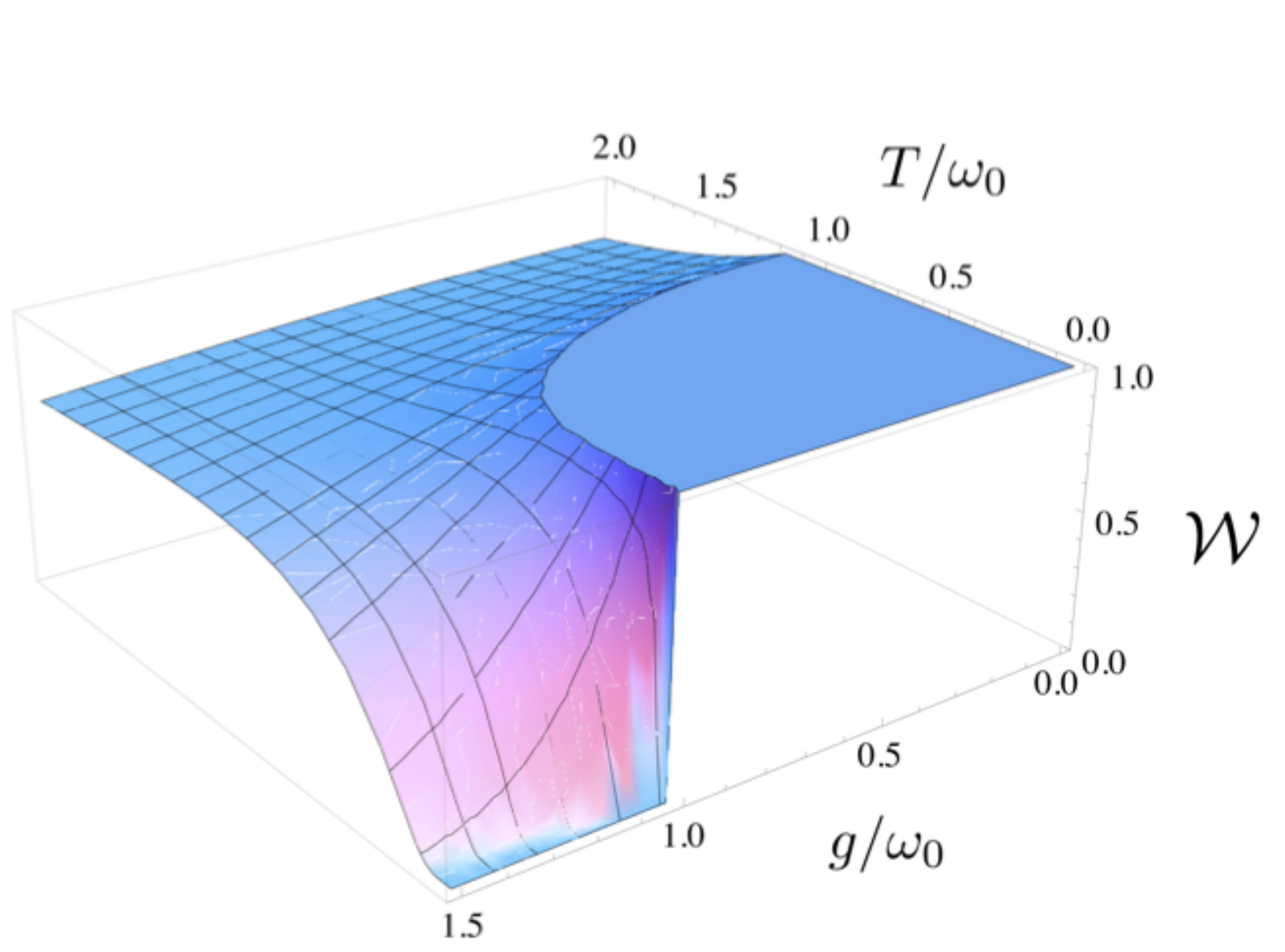}
\caption{Violation of the variance based entanglement witness ${\cal W}$ proposed in \cite{vedral}  and detailed in the text for a two--qubit system (separability threshold equal to 1) as a function of $T$ and $g$. This witness also displays increasing violation with temperature and is currently measured in solid state systems with a reduced controllability. Its  detailed experimental study as a function of $g$ and $T$ can  provide better relative violation than computing the negativity of the density matrix \cite{SI}.  It also presents the advantage of being more adapted to large scale systems, since it relies on collective measurements only.}
\label{fig4}
\end{figure}

Thermal reservoirs with different interesting properties can be engineered using  the Kraus formalism. In order to show this, we study as an example the basis transformation $\hat U=e^{-i\sum_j^N g_j\sigma_z^{(j)}}e^{-i\frac{\pi}{4}\left (\sum_{j=1}^N \hat \sigma_+^{(j)}\hat \sigma_-^{(j+1)}+\hat \sigma_-^{(j)}\hat \sigma_+^{(j+1)}\right )}$ realized in a chain on $N=2$  non interacting qubits. When applied to the jump operators $\hat \sigma_{\pm}^{(i)} (i=1,2)$, the considered transformation $\hat U$ leads to a new set of jump operators $ \hat U  \hat \sigma_-^{(i)} \hat U^{\dagger}=\hat{E}_{i} $, $ \hat U \hat \sigma_+^{(i)} \hat U^{\dagger}=\hat{E}_{i}^{\dagger} $ that create transitions between entangled states. The physical meaning of  the entangling operation $\hat U$ is that it creates an engineered reservoir where the eigenstates of the Hamiltonian  $\hat H=\sum_{j=1}^Ng/2\left ( \hat \sigma_+^{(j)}\hat \sigma_-^{(j+1)}+\hat \sigma_-^{(j)}\hat \sigma_+^{(j+1)} \right) $ are independently coupled to the environment. In the $2$ qubit case, such eigenstates  are given by: $\vert \tilde{0} \rangle \rightarrow \vert 0_1 0_2 \rangle$, $  \vert \tilde{1} \rangle \rightarrow \frac{1}{\sqrt{2}} ( \vert 0_1 1_2 \rangle + i \vert 1_1 0_2 \rangle )$, $  \vert \tilde{2} \rangle  \rightarrow  \frac{1}{\sqrt{2}} ( i \vert 0_1 1_2 \rangle + \vert 1_1 0_2 \rangle )$, $  \vert \tilde{3} \rangle \rightarrow  \vert 1_1 1_2 \rangle$. The transformed jump operators are given by $\hat E_i= (  \hat{\sigma}_{-}^{(i)} + i \, \hat{\sigma}_{z}^{(i)} \hat{\sigma}_{-}^{(j)}   ) / \sqrt{2}$, $i,j=1,2$. The transformed level scheme with corresponding energies and jump operators is displayed in Fig. \ref{fig1}(b), and can be compared to the one displayed in Fig. \ref{fig1}(a).  We can see that, in both cases, there are two independent reservoirs, each of them coupling different transitions that can involve both entangled and separable states. From the spectrum shown in Fig. \ref{fig1}(b), we can see that this particular  transformation  opens the perspective of simulating and studying thermal entanglement in a controlled and systematic way:   for $T=0$ and $g<\omega_o$, where $\omega_o$ is the transition frequency  between states $\ket{0_i}$ and $\ket{1_i}$ $(i=1,2)$, the stationary state (ground state) is separable, but entanglement appears with increasing temperature.  Fig. \ref{fig4} displays the dependency on $T$ and $g$ of a variance based entanglement witness \cite{vedral}: ${\cal W}=\sum_{\alpha=x,y,z}\Delta \left (\sum_i^N \hat \sigma_{\alpha}^{(i)}\right )\geq N/2$. This witness, that only involves collective measurements, was experimentally tested in solid state systems \cite{NMR, Tatiana}, whose violation also increases with temperature, contrary to common sense. In the same Figure, we can also observe quantum phase  transitions by changing the reservoir's properties, or making $g\leq \omega_o$. If $g\geq \omega_o$, the ground state is entangled (quantum phase transition point), and entanglement decreases with increasing temperature. In Fig. 10 of \cite{SI} we show the dependency of the negativity, necessary and sufficient criterium to detect entanglement in a bipartite qubit system, as a function of $T$ and $g$ as well. 

Such transformations can be implemented experimentally in a variety of systems such as single photons and trapped ions. A detailed implementation relying on the different degrees of freedom of a single photon is presented in \cite{SI}. Below, we discuss the use of a trapped ions system, and  show a generalization of \cite{Blatt}, leading to the observations of the $T$ and $g$ dependency of entanglement in a controllable and integrable quantum system. 

\begin{figure}[t!]
\centering
\includegraphics[width=0.4\textwidth]{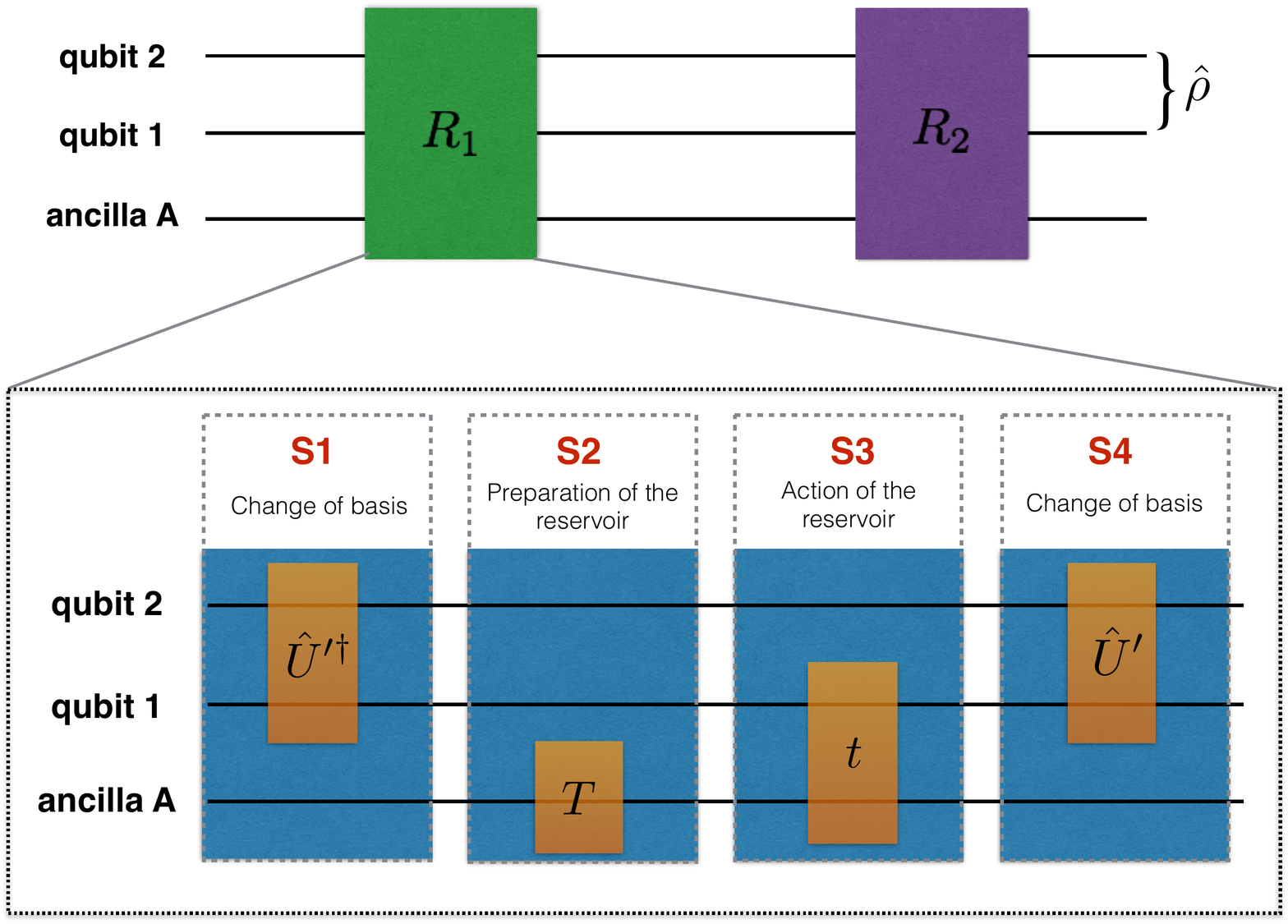}
\caption{Schematic representation of the proposed thermal reservoir engineering for a two--qubit system, involving the sequential application of  two parts, $R_1$ and $R_2$. In the detailed scheme, the sequence of operations involved in the realization of $R_1$ is shown. The sequence involved in the realization of $R2$ is exactly the same, with different choices of parameters. In $S1$, a change of base is realized in the qubit system. Then, in $S2$ the ancilla ion is prepared in a mixed state. The amount of entanglement is related to the reservoir's temperature: a separable state corresponds to  $T=0$ while a maximally entangled state to  $T\rightarrow \infty$. $S3$ simulates the action of a local thermal reservoir on qubit 1 using a SM gate. The duration of this gate, and consequently, the amount of entanglement it creates, is related to the effective time elapsed since the system was coupled to the reservoir and started to controllably ``decohere". If the gate realizes a ``$\pi$ pulse", we have the equivalent to the production of the steady state of the system ($t\rightarrow \infty$). In $S4$, the basis change is undone. After the successive application of $R_1$ and $R_2$, measurement of the two-qubit state leads to the density matrix that would have been obtained by solving (\ref{MasterEq2}) with $\hat c_i\rightarrow \hat E_i$ and $\hat c_i^{\dagger}\rightarrow \hat E_i^{\dagger}$. }
\label{fig3}
\end{figure}

The action of each transformed reservoir is independently simulated and thus the protocol is split in two parts, $R_1$ and $R_2$. In both of them, we make use of  auxiliary ions (ancillae) to mimic the role of the reservoirs. Such ancillae are prepared in judiciously chosen mixed initial states whose degree of purity  is related to the temperature of the environment.  After step $R_1$, the ancilla is reinitialized and prepared to simulate the action of $R_2$.  Both parts of the protocol follow exactly the same sequence of steps, summarized in Fig.~\ref{fig3}, differing only by the choice of parameters involved in the interactions between each pair of ions. $R_1$ starts with a  change of basis in the two qubit system (step S1), performed by the operator $\hat U'^{\dagger}=e^{i\frac{\pi}{4}\left (\sum_j^N \hat \sigma_+^{(j)}\hat \sigma_-^{(j+1)}+\hat \sigma_-^{(j)}\hat \sigma_-^{(j+1)}\right )}$ \cite{comment}. This operation can be experimentally implemented in a trapped ion system using an entangling gate, as the celebrated  gate proposed by S\o rensen and M\o lmer (SM) \cite{SM} and currently used with high fidelity \cite{Fidelity}.  With such gates, collective operations $e^{i \Omega \tau\left (\hat \sigma_{\alpha}^{(1)}+\hat \sigma_{\alpha}^{(2)}\right )^2}$ can be realized, where $\alpha$ is an arbitrary direction in the three dimensional space, $\Omega$ is proportional to the effective lasers-ions coupling and $\tau$ is the laser--ion interaction time. The parameter $\alpha$ is determined by the  relative phases of the lasers used to perform the SM gate. The advantage of realizing this basis change operation is that we can now simulate the action of a reservoir that acts locally and independently on each qubit, where quantum jumps involve the application of operators $\hat \sigma_{\pm}^{(i)}$. Then, a subsequent application of the operator $\hat U'^{\dagger} $ will undo the basis change, and the net effect is the perfect reproduction of  the action of an engineered reservoir where quantum jumps are realized by  the transformed operators $\hat E_i$, $\hat E_i^{\dagger}$. 

In a second step, S2, the  ancilla ion is prepared in state $\hat \rho_{A}^{R_1}=p_1\ket{0_A}\bra{ 0_A}+(1-p_1)\ket{1_A}\bra{ 1_A}$. Since the experiment must be repeated several times to acquire statistical data, this state can be prepared by using different ancilla states, $\ket{0_A}$ or $\ket{1_A}$ at each run of the experiment, respecting the required classical statistical distribution. This  statistical weight is related to the environment's temperature: $p_1=1$ (pure reservoir state) means $T=0$, while $p_1=1/2$ (completely mixed state) means $T\rightarrow \infty$. 

We can thus move to step S3, that simulates the cooling and heating processes. This is done using again a SM gate. However, in this step, it couples qubit $1$ to  the auxiliary ion $A$ leading to: $\ket{0_1}\ket{1_A}\rightarrow \sqrt{\lambda_1(t)}\ket{0_1}\ket{1_A}-\sqrt{1-\lambda_1(t)}\ket{1_1}\ket{0_A}$ and $\ket{1_1}\ket{0_A}\rightarrow \sqrt{1-\lambda_1(t)}\ket{1_1}\ket{0_A}+\sqrt{\lambda_1(t)}\ket{0_1}\ket{1_A}$. Notice that, in this transformation, $t$ is the time involved in the definition of the Kraus operators (see Eq.~\eqref{Kraus}), it corresponds to the  evolution time under the Lindblad equation~\eqref{MasterEq2}. It should not be confused with the interaction time $\tau$ needed to entangle the qubit and the ancilla $A$. These two parameters relate as follows: $\tau={\rm arccos}(\sqrt{\lambda_1(t)})/\Omega$.

In order to close the $R_1$ part of the protocol (S4), one should, in principle, apply $\hat U'$ to convert the system back to the original basis, as schematized in Fig. \ref{fig3}. Nevertheless, since the $R_2$ starts with S1, which is an application of operation $\hat U'^{\dagger}$, the combination of both steps is nothing but the identity operation.    

We can verify that after $R_1$, reservoir and qubits are entangled, and when tracing out the ancilla (environment) ion, one  obtains exactly the four Kraus operators associated to a thermal reservoir.  In order to realize $R_2$, one can either add another ancilla ion, playing a role analogous to the previous ones, or reinitialize the already used ions, preparing them  in a state convenient to the realization of $R_2$. 

State reinitialization can be achieved through the following sequence, experimentally realized in \cite{Blatt}: the physical state $\ket{1_A}$ encoding quantum information and used as ancilla is a  long-lived internal states. However, it can be coupled to unstable states that rapidly spontaneously decay to state $\ket{0_A}$. In this effective  incoherent process, the information encoded in the ancilla ions is transmitted, through spontaneous emission, to the ``real" macroscopic environment, so the non-unitary character of the evolution of the two-bit system is preserved, even though the  ancilla ion is  in state $\ket{0_A}$,  so not entangled to the qubits anymore.

From such a reinitialized state, we can now prepare the ancilla in state $\hat \rho_{A}^{R_2}=p_2\ket{0_A}\bra{ 0_A}+(1-p_2)\ket{1_A}\bra{ 1_A}$ using the same procedure  and follow exactly the same prescription as in  $R_1$, replacing $\lambda_1(t)$ by $\lambda_2(t)$.

After the action of $R_1$ and $R_2$, by judiciously choosing $p_j=\frac{\bar n_j+1}{2\bar n_j+1}=\left (e^{-\frac{\omega_o+(-1)^j g}{k_BT}}+1\right )^{-1}$ and $\lambda_j(t)=1-e^{-\Gamma\left (2\bar n_j+1\right)t}$, $j=1,2$ we have that the two qubit state is  prepared in a mixed state. It corresponds to the solution  of a Lindblad equation describing its coupling to an engineered reservoir with exotic, experimentally chosen, properties. 

Scaling of the presented protocol is possible, at the expense of either reinitializing ancillae a number of times that scales linearly with the number of qubits $N$ or by adding a number of  ancillae ions that also scales linearly with $N$.

In conclusion, we proposed a method to engineer quantum markovian  reservoirs at finite temperature and illustrated it by showing how to engineer a reservoir leading to a steady state displaying thermal entanglement, for some choices of parameters. Other choices of parameters and entangling transformations can be made, leading to the engineering of finite temperature reservoirs with different asymptotic properties. Finite temperature reservoir engineering in controllable and integrable quantum systems, as trapped ions, superconducting systems and single photons, opens the perspective to a deeper experimental study of the entanglement dynamics dependency with temperature, and all the related surprising and counterintuitive phenomena that remained unexploited so far. The experimental implementation of the suggested protocols is within immediate reach in different experimental set-ups using current technology.


\begin{thebibliography}{99}

\bibitem{Zurek} W. H. Zurek, Rev. Mod. Phys. {\bf 75}, 715 (2003). 

\bibitem{Poyatos} J. F. Poyatos, J. I. Cirac, and P. Zoller, Phys. Rev. Lett. {\bf 77}, 4728 (1996).


\bibitem{Carvalho}  A. R. Carvalho, P. Milman, R. L. de Matos Filho and L. Davidovich, Phys. Rev. Lett. \textbf{86}, (2001). 

\bibitem{Rouchon} A. Sarlette, Z. Leghtas, M. Brune, J. M. Raimond and P. Rouchon, Phys. Rev. A {\bf 86}, 012114 (2012).

\bibitem{Wineland} C. Myatt {\it et al.}, Nature {\bf 403}, 269 (2000). 

\bibitem{Blatt} J. T. Barreiro {\it et al.}, Nature {\bf 470}, 486 (2011). 

\bibitem{Wineland2} Y. Lin {\it et al.}, Nature {\bf 504}, 415 (2013).

\bibitem{Krauter} H. Krauter {\it et al.},  Phys. Rev. Lett. 107, 080503 (2011).


\bibitem{SC} Z. Leghtas {\it et al.},  Phys. Rev. A 88, 023849 (2013).

\bibitem{SCexp} S. Shankar {\it et al.}, Nature {\bf 504}, 419 (2013).

\bibitem{Haroche13} S. Haroche, Rev. Mod. Phys. \textbf{85}, 1083 (2013)

\bibitem{Wineland13} D. J. Wineland, Rev. Mod. Phys. \textbf{85}, 1103 (2013)

\bibitem{Steve} G. H. Aguilar {\it et al.}, Phys. Rev. Lett. {\bf 112}, 160501 (2014). 

\bibitem{Thermal1} M. C. Arnesen, S. Bose, and V. Vedral, Phys. Rev. Lett. {\bf 87},
017901 (2001).

\bibitem{Thermal2} A. Osterloh, L. Amico, G. Falci, and R. Fazio, Nature (London)
{\bf 416}, 608 (2002).

\bibitem{NMR} A. M. Souza, M. S. Reis, D. O. Soares-Pinto, I. S. Oliveira and R. S. Sarthour, Phys. Rev. B {\bf 77}, 104402 (2008). 

\bibitem{Ghosh} S. Ghosh, T. F. Rosenbaum, G. Aeppli, and S. N. Coppersmith, Nature (London) {\bf 425}, 48 (2003).
\bibitem{Vertesi} T. V\ 'ertesi and E. Bene, Phys. Rev. B {\bf 73}, 134404 (2006).
\bibitem{Brukner} C. Brukner, V. Vedral, and A. Zeilinger, Phys. Rev. A {\bf 73}, 012110
(2006).
\bibitem{Tatiana}  T. G. Rappoport, L. Ghivelder, J. C. Fernandes, R. B. Guimar$\tilde {\rm a}$es,
and M. A. Continentino, Phys. Rev. B {\bf 75}, 054422 (2007).

\bibitem{Marcelo} M. F. Santos, P. Milman, L. Davidovich and N. Zagury, Phys. Rev. A {\bf 73}, 040305 (2006). 

\bibitem{SteveSD} M. P. Almeida {\it et al.}, Science {\bf 316 }, 579 (2004).

\bibitem{Eberly} T. Yu and J. H. Eberly, Phys. Rev. Lett. {\bf 93}, 140404 (2004).

\bibitem{James} A. Al-Qasimi and D. F. V. James, Phys. Rev. A {\bf 77}, 012117 (2008). 

\bibitem{Birth} C. E. L\'opez, G. Romero, F. Lastra, E. Solano, and J. C. Retamal, Phys. Rev. Lett. {\bf 101}, 080503 (2008).



\bibitem{Transp1} A. Rivas, A.D. Plato, S. Huelga and M.B. Plenio, New J. of Phys.  {\bf 12}, 113032 (2010).  
\bibitem{Transp2} P. Rebentrost, M. Mohseni, I. Kassal, S. Lloyd and A. Aspuru-Guzik, New J. of Phys.  {\bf 11}, 033003 (2009).  

\bibitem{Thermo} R. Dorner, J. Goold, C. Cormick, M. Paternostro, and V. Vedral, Phys. Rev. Lett. {\bf 109}, 160601 (2012).

\bibitem{vedral} M. Wiesniak, V. Vedral and C. Brukner, N. Journ. Phys. {\bf 7}, 258 (2005). 

\bibitem{SI} Supplemental Material

\bibitem{Lindblad}  H.-P. Breuer and F. Petruccione , \textit{The Theory Of Open Quantum Systems} Oxford University Press (2002). 

\bibitem{Castin} K. M\o lmer, Y. Castin and J. Dalibard, J. Opt. Soc. Am. B {\bf 10}, 524 (1993). 

\bibitem{Buch} S. Sauer, C. Gneiting and A. Buchleitner, Phys. Rev. A {\bf 89}, 022327 (2014). 

\bibitem{Chuang} M. Nielsen and I. Chuang {\it Quantum Information and Quantum Computation}, Cambridge University Press, (2004). 
 
 \bibitem{Squeeze} R. Srikanth and S. Banerjee, Phys. Rev. A {\bf 77}, 012318 (2008). 


\bibitem{comment} The absence in $\hat U'$ of the qubit dependent dephasing factor $e^{i\sum_i g_i \hat \sigma_z^{(i)}}$ appearing in $\hat U$  is due to the following: this factor leads to a breaking of degeneracy of the qubits. In the discussed reservoir engineering procedure, such symmetry breaking can be done by experimentally controlling the coupling between the ions encoding the qubits and the auxiliary ions used to simulate the action of the reservoir.  This is related to the dependency of $p_i$$i=1,2$ with the energy difference $g$ between the two entangled states $\ket{\tilde 1}$ and $\ket{\tilde 2}$of the transformed basis. 

\bibitem{SM} A. S\o rensen and K. M\o lmer, Phys. Rev. Lett. {\bf 82}, 1971 (1999). 

\bibitem{Fidelity} J. Benhelm, G. Kirchmair, C. F. Roos and R. Blatt, Nat. Phys.{\bf 4}, 463 (2008).




\end{thebibliography}
\end{document}